\documentstyle[prb,amsfonts,twocolumn,aps,epsfig]{revtex}
\begin{document}
\newcommand{\SiSj}{\vec{S}_i\cdot\vec{S}_j}
\title{Spin-Liquid phase of the Multiple-Spin Exchange Hamiltonian
on the Triangular Lattice}
\author{
G. Misguich  
\thanks{Laboratoire de Physique Th{\'e}orique des
	Liquides, Universit{\'e} P. et M. Curie, case 121, 4 Place Jussieu,
	75252 Paris Cedex. UMR 7600 of CNRS.},
C. Lhuillier  $^*$,
B. Bernu  $^*$
and C. Waldtmann
\thanks{Institut f{\"u}r Theoretische Physik,
Universit{\"a}t Hannover, D-30167 Hannover, Germany}
}
\date{\today}

\maketitle

\bibliographystyle{prsty}

\begin{abstract}
We performed an exact diagonalization study of the spin liquid phase
of the Multiple-Spin Exchange model on the triangular
lattice. It is characterized
by no N{\'e}el Long Range Order (NLRO), short-ranged magnetic correlations
and a spin gap. We found no LRO in any local order parameter we investigated
(chiral, dimer,...).  The probable asymptotic ground state degeneracy is
discussed.
We argue that it could be of topological origin and that the system is
probably not a chiral spin liquid. A possible relation to the
Affleck-Kennedy-Lieb-Tasaki (AKLT) phase is discussed.

\end{abstract}


\begin{section}{Introduction}

{\it A general model}. ---
The Multiple Spin Exchange (MSE) model was introduced by Thouless \cite{t65}.
It is an effective description of the magnetic properties
of quasi-localized fermions. 
This lattice spin Hamiltonian was introduced
to describe the magnetic consequences of tunneling events where several
indistinguishable
particles exchange their positions. For spin-$\frac{1}{2}$ particles,
when two-body processes are the
only important events, it leads to the familiar Heisenberg Hamiltonian:
\begin{equation}
	H=\sum_{i<j} J_{i,j}P_{i,j}=\sum_{i<j} J_{i,j}
	\left(2\SiSj+\frac{1}{2}\right)
\end{equation}
where $P_{i,j}$ is the spin permutation operator, and $2J_{i,j}$ the
(positive) frequency of the tunnel process exchanging the particle on site $i$
and the particle on site $j$.
But in low density crystals, cyclic
exchange processes $P_n$ of $n\geq 3$ particles can be important.
It is the case in low density solid $^3$He films (\cite{rbbcg98}
and references therein) and
in the Wigner crystal\cite{r84}.
More generally, multiple-spin interactions
with $n \geq 2$ are expected in the spin systems
where interactions are strong and cannot be reduced to Heisenberg couplings.
In numerous quantum crystals,
such couplings are expected to be generated by integrating out the
non-magnetic degrees of freedom. For instance, the role played by
MSE in the context of metal-insulator transitions has been discussed
recently by Chakravarty {\it et al.}\cite{cknv98}.  Therefore it is of wide
interest to understand the nature of the magnetic fluctuations introduced
by MSE couplings.

{\it The MSE spin liquid phase}. ---
In a previous work we studied the phase diagram
of the MSE model on the triangular lattice and showed the existence
of a spin liquid phase \cite{mblw98}.
In this work we present a detailed characterization
of the low energy physics of this new spin liquid. The hamiltonian
involves the five simplest ring exchange patterns on the
triangular lattice:

\begin{eqnarray}\label{Hmulti}
H=J_2 \sum_{
\begin{picture}(17,10)(-2,-2)
	\put (0,0) {\line (1,0) {12}}
	\put (0,0) {\circle*{5}}
	\put (12,0) {\circle*{5}}
\end{picture}
} P_2
-J_3 \sum_{
\begin{picture}(26,15)(-2,-2)
        \put (0,0) {\line (1,0) {12}}
        \put (0,0) {\line (3,5) {6}}
	\put (12,0) {\line (-3,5) {6}}
        \put (6,10) {\circle*{5}}
        \put (0,0) {\circle*{5}}
        \put (12,0) {\circle*{5}}
\end{picture}
} \left( P_3+P_3^{-1}\right) \nonumber \\
+J_4 \sum_{
\begin{picture}(26,15)(-2,-2)
        \put (0,0) {\line (1,0) {12}}
        \put (6,10) {\line (1,0) {12}}
        \put (0,0) {\line (3,5) {6}}
        \put (12,0) {\line (3,5) {6}}
        \put (6,10) {\circle*{5}}
        \put (18,10) {\circle*{5}}
        \put (0,0) {\circle*{5}}
        \put (12,0) {\circle*{5}}
\end{picture}
} \left( P_4+P_4^{-1}\right)
-J_5 \sum_{
\begin{picture}(26,15)(-2,-2)
        \put (0,0) {\line (1,0) {24}}
        \put (6,10) {\line (1,0) {12}}
        \put (0,0) {\line (3,5) {6}}
        \put (18,10) {\line (3,-5) {6}}
        \put (6,10) {\circle*{5}}
        \put (18,10) {\circle*{5}}
        \put (0,0) {\circle*{5}}
        \put (12,0) {\circle*{5}}
        \put (24,0) {\circle*{5}}
\end{picture}
} \left( P_5+P_5^{-1}\right)  \\
+J_6 \sum_{
\begin{picture}(26,30)(-2,-15)
        \put (6,10) {\line (1,0) {12}}
        \put (6,-10) {\line (1,0) {12}}
        \put (0,0) {\line (3,5) {6}}
        \put (0,0) {\line (3,-5) {6}}
        \put (18,10) {\line (3,-5) {6}}
        \put (18,-10) {\line (3,5) {6}}
        \put (6,10) {\circle*{5}}
        \put (6,-10) {\circle*{5}}
        \put (18,10) {\circle*{5}}
        \put (18,-10) {\circle*{5}}
        \put (0,0) {\circle*{5}}
        \put (12,0) {\circle*{2}}
        \put (24,0) {\circle*{5}}
\end{picture}
} \left( P_6+P_6^{-1}\right) \nonumber
\end{eqnarray}

Due to the Pauli principle, exchange of an odd number of fermions is
ferromagnetic whereas exchange
of an even number if antiferromagnetic. This is the reason for the
alternating signs in Eq. (\ref{Hmulti}) (all the $J_n$ being $>0$).
Because of the equality $P_{1,2,3}+P_{3,2,1}=P_{1,2}+P_{2,3}+P_{3,1}-1$,
valid for spin-$\frac{1}{2}$, triple exchange around triangles
can be taken in account just by modifying the bare $J_2$ frequency
$J_2\to J_2-2J_3$. In the following we thus assume without
any loss of generality that $J_3=0$
and that $J_2$ can be $\geq0$ or $<0$.

At $T=0$ there is a first order phase transition between a paramagnetic
phase (the ground state is a singlet $S=0$) and a ferromagnetic one
(the ground state is fully polarized: $S=N/2$) \cite{mblw98}. In this work
we are interested in the ground state and first excited states
structure in the paramagnetic phase.
The behavior of this spin liquid when an external magnetic field is applied
will be discussed elsewhere.

{\it What kind of Spin Liquid is the MSE model ?} ---

Many different quantum ground states, 
which differ by their spatial broken symmetries and excitation spectra,
have short-ranged spin-spin correlations and an $S=0$
ground state. Among the two-dimensional Hamiltonians
with a single spin in the unit cell (no dimerization nor inequivalent bonds)
there are three important examples where the ground state wave-function is
understood:
\begin{itemize}
	\item Spin$-\frac{1}{2}$ Klein\cite{k82} models.
	This family of Hamiltonians
	generalizes
	the Majumdar-Gosh\cite{mg69} model. Since any first-neighbor
	valence-bond state is a ground state, the degeneracy is extensive
	in two dimensions.
	\item Valence Bond Crystal:
	It has some dimer-dimer long-ranged order or some
	more complicated plaquette order.
	It has a spin gap and spin-spin correlations
	are short-ranged.
	The translation symmetry is spontaneously broken and therefore
	the ground state is degenerate.
	An example may be the frustrated antiferromagnet on the square
	lattice
	\cite{sz92,zu96}. These spatial symmetry
	breakings are found in large-$N$ limits of $SU(N)$
	models\cite{rs89,rs90,rs91}
	when the
	spin $S$ does not match the coordination number $z$ ($2S\neq0$~mod~$z$).
	
	\item Valence Bond Solid (Affleck {\it et al.}\cite{aklt87}).
	It exists when the spin $S$ on one site
	matches the $z$ coordination number of the lattice
	($2S=z$). It breaks no symmetry and has a spin gap but, a priori, it
	cannot be constructed with spin-$\frac{1}{2}$.
	The Haldane phase\cite{h88}
	in one dimension (integer spin) belongs to the same universality
	class. The ground state is non-degenerate (infinite volume limit
	with periodic boundary conditions),
	spin-spin and dimer-dimer correlations decay exponentially
	with distance.
\end{itemize}

There are very few models where
resonances between short-ranged valence-bond states select a single
(and fully symmetric under symmetry operations)
combination of them. A perturbed
Klein Hamiltonian could be a realization\cite{cck89}.
The $J_1$-$J_2$-$J_3$ model on the square lattice might also be a candidate
(exact diagonalizations on a $N=16$ sample \cite{fkksrr89}).
So far, the short-ranged Resonating Valence-Bond (RVB) picture proposed
by Anderson\cite{a73} has not yet found an explicit realization:
there is no definite spin-$\frac{1}{2}$ Hamiltonian in two dimensions with 
no broken translation symmetry and a non degenerate ground state.
From this point of view the MSE is of great relevance since our numerical
data point to short-ranged correlations with no kind of LRO.

\end{section}
\begin{section}{Range of the spin Liquid phase in the MSE model
		on the triangular lattice}

Even in the classical limit, the ground state
of the MSE Hamiltonian is exactly known only in three limits:
\begin {itemize}
	\item pure $J_2>0$ case: 3-sublattice N{\'e}el State
	\item pure $J_4$ case: 4-sublattice N{\'e}el State
(tetrahedral state found by Momoi, Kubo and Niki\cite{km97,mkn97}).
	\item pure $J_n$ with odd n: Ferromagnetic.
\end{itemize}

An approximate phase diagram for the classical system
at $T=0$ has been obtained in the variational subspace
of planar helices \cite{mbl98}, and in the 4-sublattice subspace
for the $J_2-J_4$ model (Kubo {\it et al}. \cite{km97}).
From these results we can sketch
the {\em qualitative} shape of the classical diagram
(dotted line of Fig \ref{diag}).

\vspace{.4cm}
In the $S=\frac{1}{2}$ quantum system there is hardly any exact
result, but some regions of the diagram are understood:

\begin{itemize}
	\item Pure $J_2>0$ case: the A.F. Heisenberg Hamiltonian
	has 3-sublattice N{\'e}el RLO \cite{blp92}.
	\item Large region about the pure $J_4$ case: no
	Long Range Order (LRO),
	finite spin gap and short-ranged spin-spin
	correlations \cite{mblw98}.
	\item Large ferromagnetic phase including the pure $J_2<0$
	and pure $J_5$ Hamiltonians.
\end{itemize}

A possibility for the quantum phase diagram is presented Fig. \ref{diag}.
This simple guess relies on the following data:
\begin{itemize}
	\item Preliminary exact diagonalizations
	results indicate that the extension of
	3-sublattice N{\'e}el phase in the $J_2-J_4$ model
	is strongly reduced by 4-spin exchange. This is already the case
	at the mean field Schwinger Boson level \cite{mbl98}. The
	possibility that this N{\'e}el phase is destroyed by an infinitely small
	$J_4$ is not excluded (additional work is in progress).
	\item Our exact diagonalization data show no LRO
	in the large-$J_4$ region.
	\item The energy of the ferromagnetic state is the same
	in the classical
	and quantum cases but A.F. configurations gain energy from
	quantum fluctuations.
	Therefore, the ferromagnetic phase is slightly
	reduced in the $S=\frac{1}{2}$ case.
\end{itemize}

	On the $J_2(>0)$ - $J_5$ line, we suggest the existence of a spin
	liquid window
	between the ferromagnetic and the 3-sublattice N{\'e}el phase.
	However, this has not yet been numerically investigated.
	Another possibility is a first order phase transition
	between NLRO and ferromagnetism.
\begin{figure}
	\begin{center}
	\mbox{\psfig{figure=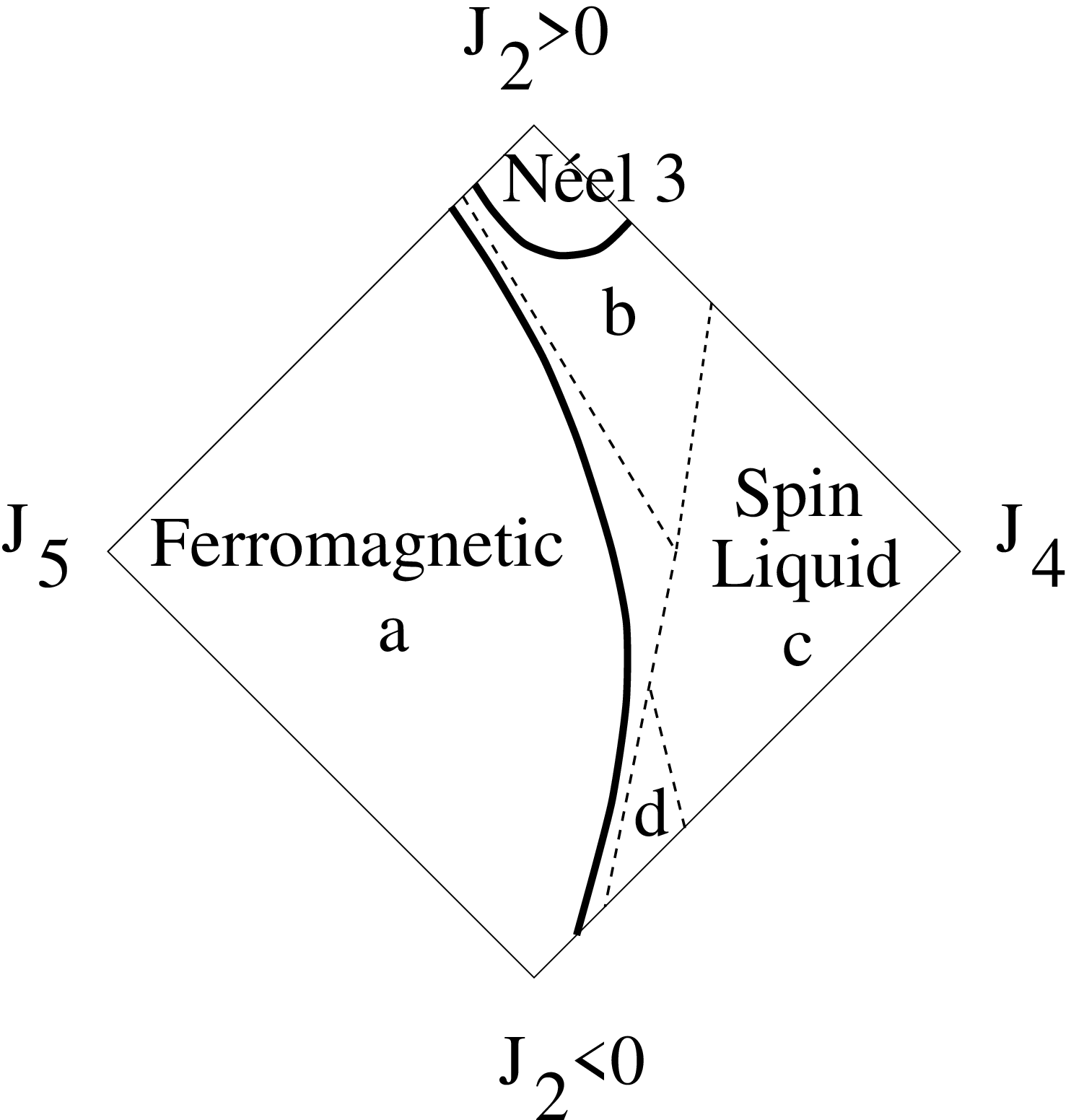,width=6cm}}
	\end{center}
	\caption[99] {Qualitative phase diagram of the MSE model
	on the triangular lattice.
	The right corner is the pure $J_4$ model, the left one
	is the pure $J_5$ one.
	Down and up corners are the ferromagnetic and antiferromagnetic
	Heisenberg Hamiltonians.
	{\bf Letters and dashed lines}: Schematic phase diagram for
	the classical model. {\bf a}: Ferromagnetic, {\bf b}:
	3-sublattice N{\'e}el
	state, {\bf c}: 4-sublattice N{\'e}el state, {\bf d}: long wave length
	spiral state, or ferrimagnetic state. 
	{\bf Full lines}: likely scenario for the $S=\frac{1}{2}$ quantum
phase.
	We put a small disordered window on the upper $J_4=0$ line
	but a first order transition
	between ferromagnetism and 3-sublattice N{\'e}el order is also
	possible, some more numerical work is needed to fix this point.
	}
	\label{diag}
\end{figure}

\end{section}

\begin{section}{Short-ranged magnetic correlations}
	
\begin{subsection}{no N{\'e}el Long Range Order (NLRO)}
	The first question to address in the non-ferromagnetic region is:
	Is the system N{\'e}el Long Range Ordered at $T=0$ ?
	
{\it Periodic boundary conditions}. ---
	As a N{\'e}el state breaks the $SU(2)$ and some spatial
	symmetries of $H$,
	it cannot be an exact eigenstate on a finite size system.
	It has to be a linear combination of eigenstates belonging
	to different Irreducible Representations (IR) of the spatial
	symmetry group and to different $S$
	sectors \cite{blp92,lblps97}. As the dynamics of a N{\'e}el
	order parameter
	is the one of a free rotator, the corresponding low energy levels
	scale as:
\begin{equation}
	E\simeq\frac{S(S+1)}{N\chi_0}
	\label{rotator}
\end{equation} since the inertia of that
	rotator is proportional to the number of sites $N$
	($\chi_0$ is the susceptibility per site at zero field:
	$\chi_0=\frac{1}{N}\frac{\partial <2S>}{\partial B}_{|B=0}$).
	At fixed $S$, these
	states collapse to the ground state in thermodynamic limit
	more rapidly that the softest magnon\footnote{
		The softest magnon has a total energy
		$E_{min}=c|{\bf k}_{min}|$ where
		$c$ is the slowest spin-wave velocity and
		${\bf k}_{min}\sim \frac{1}{\sqrt{N}}$ is the smallest
		wave vector allowed by the periodic boundary conditions
		on the finite size sample.
		This gives $E_{min}\simeq \frac{c}{\sqrt{N}}$.}.
	These states form a {\em  tower of states}.
	This low energy structure is absent in the spectra
	as soon as $J_4$ is not
	negligible. Fig.~\ref{spectres_N=27a},\ref{spectres_N=27b}
	shows how the
	3-sublattice N{\'e}el tower of state is destroyed by 4-spin exchange.
\begin{figure}
	\begin{center}
	\mbox{\psfig{figure=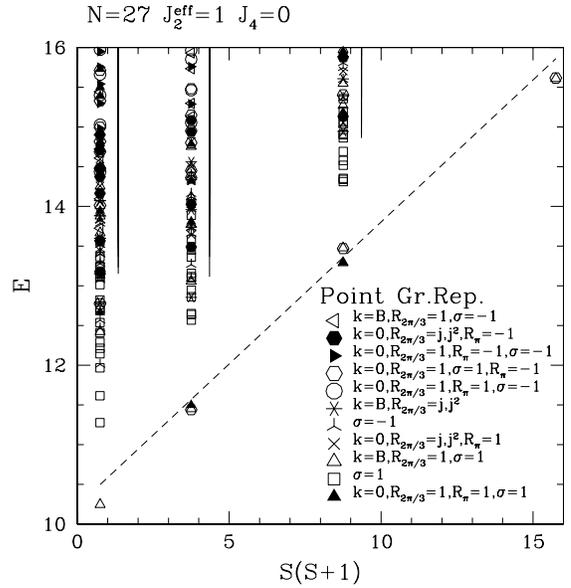,width=8cm}}
	\caption[99]{First-neighbor Heisenberg
	spectrum on the triangular lattice. The $SU(2)$ and spatial
	symmetry breakings due to the 3-sublattice Long Range N{\'e}el
	Order appear as a set of states with an energy scaling as
	$E(S)\sim S(S+1)/N$ (dashed line).
	The symbols represent the quantum number of the different
	eigenstates. ${\bf k}$ is the wave vector. ${\bf k}=B$ is
	the corner of the Brillouin zone (the impulsion is not indicated
	if ${\bf k}\neq0$ and  $\neq B$). $R_{\theta}$ is the phase factor
	obtained in a $\theta$-rotation about the origin and $\sigma$
	stands for a reflection about an axis.
	The vertical lines indicate the energy range
	where the density of
	states is high and all the eigenvalues have not	
	been computed. This has no consequence
	on the low energy part of each irreducible representation
	of the symmetry group, where the eigenstates are known
	exactly.}
	\label{spectres_N=27a}
	\end{center}
\end{figure}
\begin{figure}
	\begin{center}
	\mbox{\psfig{figure=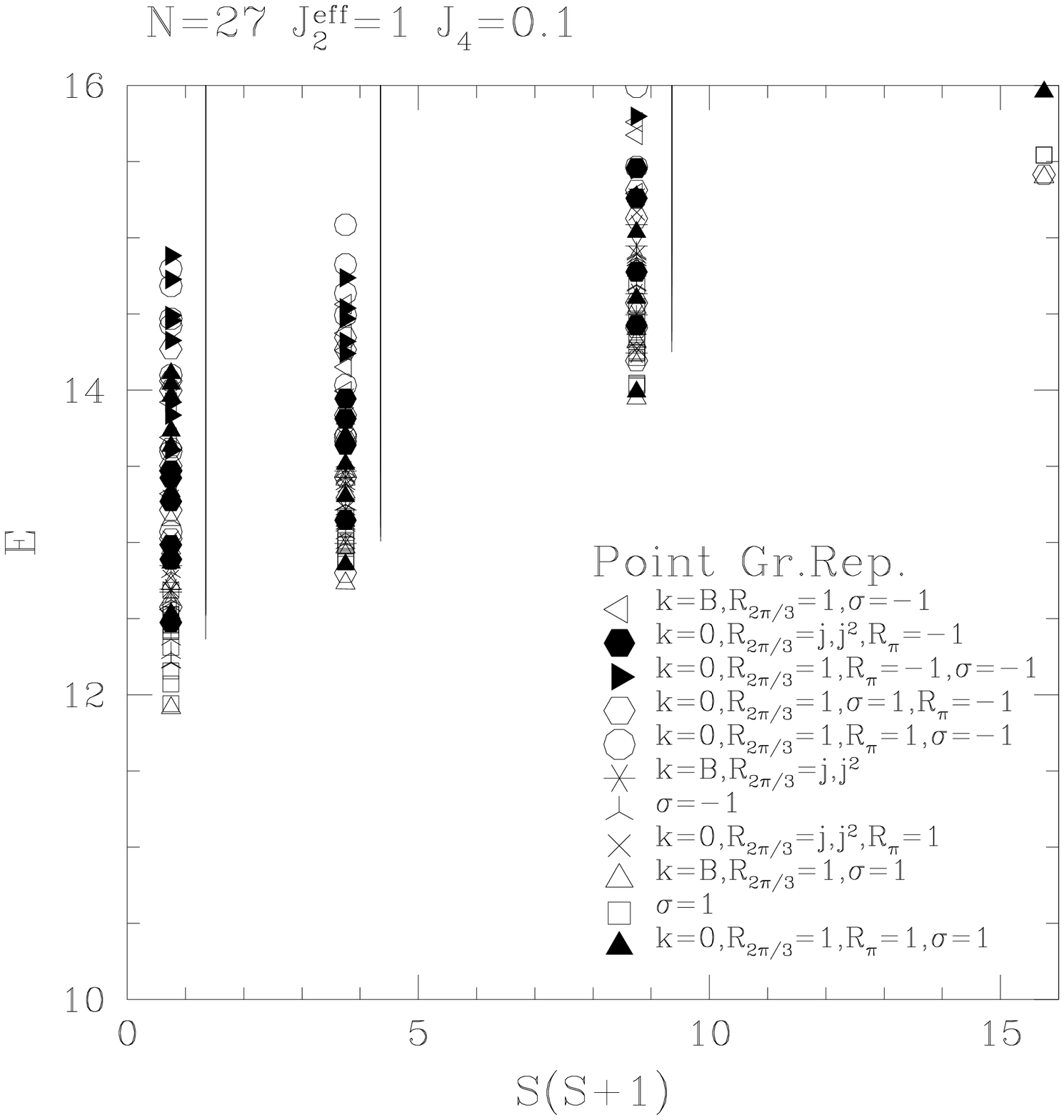,width=8cm}}
	\caption[99]{Heisenberg spectrum perturbed with 4-body
	exchange $J_4/J_2=0.1$.
	The N{\'e}el structure is destroyed: the states
	formerly embedding the symmetry breakings in the $J_4=0$
	case have been
	``pushed up'' in energy to the continuum of excitations.}
	\label{spectres_N=27b}
	\end{center}
\end{figure}
	
{\it Twisted boundary conditions}. ---
	For the tower of states to appear on a finite size sample,
	the sublattice
	structure must be compatible with the boundary conditions.
	An order with
	a long wavelength helix, or an incommensurate phase
	is therefore difficult
	to detect on small samples with periodic boundary conditions.
	Fortunately, twisted boundary conditions\footnote{
		With periodic boundary condition, the spin $\vec{S}_i$
		at site ${\bf i}$
		interacts with the spin $\vec{S}_j$ at site ${\bf j}$.
		Using the wave vector  ${\bf q}$ to twist the system
		makes the rotated spin
		${\mathcal R}_{{\bf q}\cdot{\bf i}}^z
			\left[\vec{S}_i\right]$
		interact with
		${\mathcal R}_{{\bf q}\cdot{\bf j}}^z
			\left[\vec{S}_j\right]$.
		${\mathcal R}_\theta^z$ is the rotation of angle $\theta$
		about the quantification axis~$z$. Such an Hamiltonian
		is no longer $SU(2)$ invariant when ${\bf q}\neq 0$:
		${\vec{S}_{tot}}^2=\left(\sum_i{\vec{S}_i}\right)^2$
		is no longer a conserved quantity but $S_{tot}^z$
		commutes with the twisted Hamiltonian. 
	} allow
	to overcome this
	difficulty~\cite{blp92}.
	For $N=19$ and $J_2=-2,J_4=1$, we scanned the
	Brillouin zone to determine the twist ${\bf q}_0$ which minimizes the
	ground state energy. The twist vector
	${\bf q}_0$ lies inside the Brillouin zone and has no particular
	symmetry.
	If the system had some NLRO (commensurate or not),
	this  ${\bf q}_0$ would indicate the ordering wave vector
	and the spectrum
	would show a tower of states $E\simeq\frac{(S^z)^2}{N\chi_{\perp}}$.
	Fig.~\ref{spectre_N=19}
	shows that it is not the case.
\begin{figure}
	\begin{center}
	\mbox{\psfig{figure=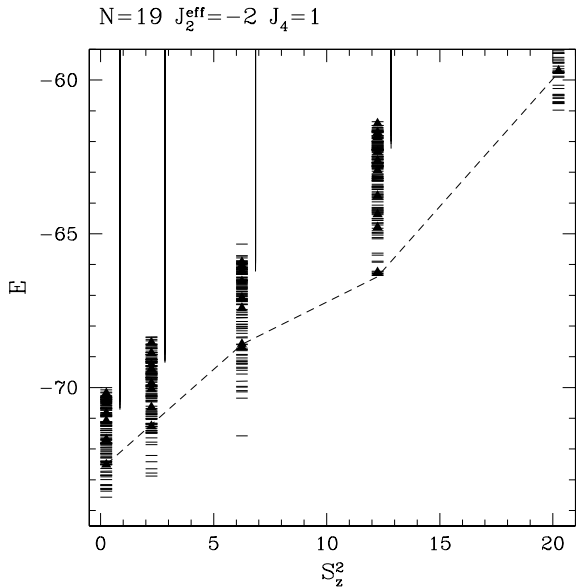,width=8cm}}
	\end{center}
	\caption[99]{Spectrum with the twisted boundary conditions
	which minimize the ground state energy.
	Eigenstates
	represented by black triangles are ${\bf k}={\bf 0}$ states.
	The spectrum shows no long-ranged (spiral) order:
	if there was a {\em tower of states} the lowest ${\bf k}={\bf 0}$
	states (joined by the dashed line) would be lower than the other
	excitations and their energies would be proportional
	to $S_z^2$. Notice that with these twisted boundary conditions
	the spatial symmetries (except translations) as well
	as the $SU(2)$ symmetry are lost.
	But $S_z$ is still a conserved quantity.
	The twist vector is ${\bf q}_0=0.27{\bf A}_1-0.20{\bf A}_2$, where
	${\bf A}_1$ and ${\bf A}_2$ are middle of the boundary of the
	Brillouin zone. By symmetry, there are twelve twists equivalent to
	${\bf q}_0$.}
	\label{spectre_N=19}
\end{figure}	
\end{subsection}

\begin{subsection}{Spin-spin correlations:
	$< {\vec{S}}_i\cdot {\vec{S}}_j > $}

For spin-$\frac{1}{2}$, $\SiSj$ varies between $-3/4$ (singlet) and $1/4$
(triplet) and the statistical average is zero.
Compared to these extremal values, the spin-spin
correlations measured in the ground state of $J_2=-2$ $J_4=1$
for $N=16,24$ and $28$ are small (Table \ref{correl}).
Fig.~\ref{fcorrel} displays the absolute value of
the spin-spin correlation as a function of distance.
On these three samples the available distances are rather small
($|i-j|\leq 3$)
and the data show important size and
orientation dependence (for $N=24$, the system has no
${\mathcal R}_{\frac{2\pi}{3}}$ symmetry and there are
3 non-equivalent directions).

\begin{table}
	\begin{tabular}{|c||c|c|c|}
$|i-j|$	& N=28 & N=24 & N=16  \\
\hline
1	&-0.06941&-0.08925 -0.06356 -0.04894            &-0.06614 \\
$\sqrt{3}$&-0.08014&-0.11194 -0.03425 -0.01640            &-0.02887 \\
2     &-0.02534&\underline{-0.17823} -0.02560 +0.02051&\underline{-0.05996}\\
$\sqrt{7}$
&\underline{+0.04983}&+0.07542 +0.00306 \underline{-0.00454}&$\times$ \\
3	  & $\times$&\underline{+0.01471} 	          &$\times$  \\
	\end{tabular}
	\caption[99]{Correlations $<\SiSj>$
	in the MSE ground state of $N=16,24$ and 28 samples
	($J_2=-2$,$J_4=1$).
	These data are plotted in Fig. \ref{fcorrel}.
	The $N=24$ sample is a $6\times4$ lattice
	which has not the ${\mathcal R}_{\frac{2\pi}{3}}$ symmetry.
	The three directions are thus non equivalent, this explains
	why $<\SiSj>$ has three possible values at distance 1.
	The underlined values are over-correlated due to periodic
	boundary conditions: they correspond to antipodal sites
	on the torus.
	}
	\label{correl} 
\end{table}

\begin{figure}
	\begin{center}
	\mbox{\psfig{figure=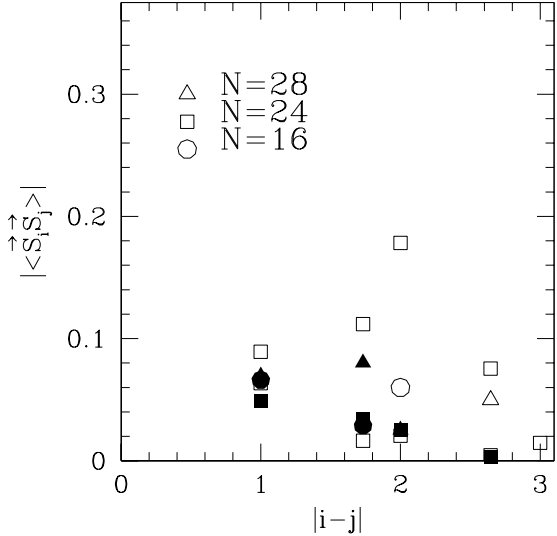,width=8cm}}
	\end{center}
	\caption[99]{Absolute value of the spin-spin correlation
	(data of Table \ref{correl}). 
	The vertical range goes from
	0 to $\frac{3}{8}$. $-\frac{3}{8}$ is the first-neighbor correlation
	in the Majumdar-Gosh chain. This scales has been chosen
	to show how small the correlations are in the MSE system.
	The filled symbols
	correspond to the correlations for which finite size effects are
	expected to be the smallest (see text).}
	\label{fcorrel}
\end{figure}

At this point, it is difficult to draw conclusions
on the spin-spin correlation decay from Fig.\ref{fcorrel}.
However, a finite size system with periodic boundary conditions
is a torus and we expect some geometrical effects for
pairs $(i,j)$ of sites which have particular symmetries.
The strongest effect is the enhancement of $<\SiSj>$
on samples where ${\bf j}-{\bf i}$ and ${\bf i}-{\bf j}$
are equivalent vectors:
antipodal sites are over correlated (underlined values
in Table \ref{correl}).
If the sample has no ${\mathcal R}_{\frac{2\pi}{3}}$ symmetry
the finite size effect on $<\SiSj>$ at distance $d=|i-j|$
should be weaker
in the direction of the vector ${\bf v}={\bf i}-{\bf j}$ which is the
not frustrated by the periodicity of the torus\footnote{
	A simple way to evaluate how the
	the ${\bf v}$ step is ``frustrated''
	by the periodic boundary conditions
	of the torus is to consider the closed loop starting from the
	origin:
	$O,O+{\bf v},O+2{\bf v},\cdots,O+(n-1){\bf v},O$. We associate
	frustration
	with the {\em winding number} of this closed curve. The minimum
	number is one:
	by successive translation of ${\bf v}$ we are back to the
	origin after winding one time around the torus.
	We make the assumption that unphysical effects due to finite
	size of the torus
	are smaller in the direction where the winding number
	is the smallest.
	Given the distance $d$, the relevant $<\SiSj>$ is thus obtained
	in the direction which gives the topologically simplest loop.
	}.
If we eliminate antipodal sites and frustrated directions
in $N=24$
only the filled symbols of Fig. \ref{fcorrel} remain. The behavior looks more
regular and the rapid decay suggests a rather short correlation length.

The point is now to understand the local structure of the ground state
wave function. There are two naive ways to build an $S=0$ state out
of a large number of spins-$\frac{1}{2}$: {\bf 1)} combine
ferromagnetically the spins in a small number of sublattices (NLRO)
and let these large spins screen themselves. {\bf 2)} combine
the spins two by two in singlets. The absence of tower of states
and significant ferromagnetic correlations have excluded
the first possibility (no NLRO).  It is easy to check that the weakness of
the A.F. correlations
makes the second scenario unlikely: the screening of a single spin at
the origin involves an important number of neighbors, up to distance
$d\simeq 2 \cdots 3$
(This is easily checked from the data of Tab.~\ref{correl}).
This will be confirmed in the next paragraph
where we show that dimer-dimer correlations are weak.

\end{subsection}
\begin{subsection}{Dimer-Dimer correlations}	

We define the dimer operator on a pair
of sites $(i,j)$ by $d_{i,j}=\frac{1-P_{i,j}}{2}$.
This projector
is 1 on a singlet and 0 on a triplet.
The dimer correlation between a reference bond $(1,2)$
and $(i,j)$ is
$D_{i,j}=< \Psi | d_{1,2} d_{i,j}|\Psi>-<\Psi|d_{1,2}|\Psi><\Psi|d_{i,j}|\Psi>$.
\footnote{
	If the system is invariant under rotation of $2\pi/3$
	$<\Psi|d_{i,j}|\Psi>$ only depends on the bond length $|j-i|$.
	}
As for the normalization, we look for the maximum value of
$D_{i,j}$. It is achieved when the two bonds are
completely correlated and gives
$D_{i,j}=<\Psi|d_{1,2}|\Psi>-<\Psi|d_{i,j}|\Psi><\Psi|d_{1,2}|\Psi>$.
So, we measure
dimer correlations by:
\begin{eqnarray}
	p_{i,j}&=&\frac{D_{i,j}}{<\Psi|d_{1,2}|\Psi>(1-<\Psi|d_{i,j}|\Psi>)}
	\\ \nonumber
	&=&\frac{<\Psi|d_{1,2} d_{i,j}|\Psi>-<\Psi|d_{i,j}|\Psi><\Psi|d_{1,2}|\Psi>}
			{\left(1-<\Psi|d_{i,j}|\Psi>\right)<\Psi|d_{1,2}|\Psi>}
\end{eqnarray}
$p_{i,j}$ is represented in
Fig.~\ref{dimera} and~\ref{dimerb}. For this quantity,
zero means that the
presence of a singlet on $(1,2)$ and the presence of a singlet on $(i,j)$
are uncorrelated. A value of one means that if a singlet exists
on $(1,2)$ there is always one on $(i,j)$.
The minimal possible value is
$p_{i,j}^{\rm min}=-\frac{<d_{i,j}>}{1-<d_{i,j}>}$,
which is $p_{i,j}^{\rm min}=-0.469$ at distance
1 in the $N=28$ ground state.
We observe negative values on the
4 bonds which are at distance 1
from the reference bond. Compared to $p_{i,j}^{\rm min}$,
this values are important (30\% ) and increase with the system size. 
This testifies an important short-ranged repulsion of parallel dimers
and excludes the possibility of spin-Peierls like structures
in which the lattice would be regularly tiled with parallel dimers.
Fig.~\ref{dimerb} shows
that the local dimer distribution favors angle of $\pm\frac{\pi}{3}$
between dimers bonds. This tendency to a $\pm\frac{\pi}{3}$ ordering of dimers
already appears in the 6-site system with open boundary
conditions. This system (shape of a triangle) 
is the smallest system with a low ground state energy and a significant
spin gap.
In larger systems, this local $\pm\frac{\pi}{3}$
structure is all the more pronounced as the ground state energy is low.

\begin{figure}
	\begin{center}
	\mbox{\psfig{figure=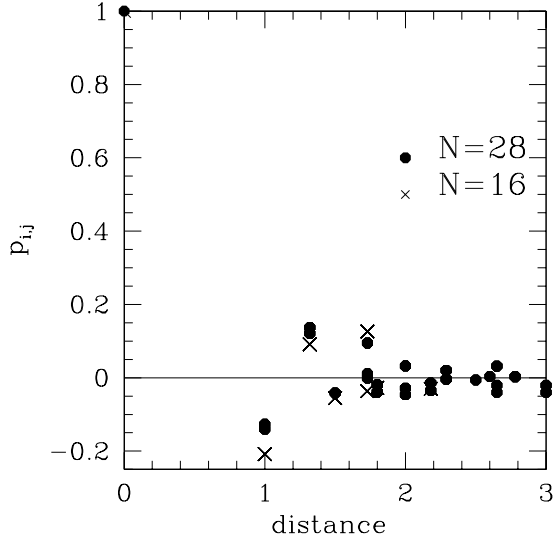,width=8cm}}
	\caption[99] {Dimer-dimer correlations in the
	ground state $|\Psi>$ of
	the MSE exchange Hamiltonian in the
	Spin-Liquid phase ($J_2=-2$ and $J_4=1$). They are plotted as
	a function of distance between bonds for $N=16$ and $28$.} 
	\label{dimera}
	\end{center}
\end{figure}

\begin{figure}
	\begin{center}
	\mbox{\psfig{figure=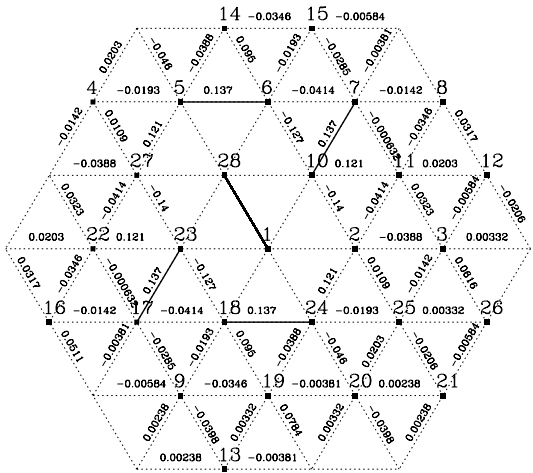,width=8cm}}
	\caption[99]{Values of $p_{i,j}$ on the $N=28$ lattice: the reference
	bond is $(1,28)$ and the 4 most strongly correlated bonds are
	in full line, they are at distance $d=\sqrt{7}/2$ from $(1,28)$
	and form a triangular pattern. We notice that the 4
	``first-neighbors''
	of the bond $(1,28)$ are strongly anti-correlated
	(see values on $(6,10)$ $(10,2)$ $(23,27)$ and $(18,23)$).}
	\label{dimerb}
	\end{center}
\end{figure}

From Fig.~\ref{dimera} one could doubt that
the dimer-dimer goes to zero with distance. However, we think it is the case
for the following reasons: {\bf 1)} dimer-dimer correlations are very weak.
Even at short distances,
the wave function has to be seen as a gaz or a liquid
and not as solid of dimers. 
{\bf 2)} The correlation at distances $d\geq 2$ decreases in a significant way
from $N=24$ to $N=28$. {\bf 3)} It is not possible to tile the triangular
lattice with dimers at $\pm\frac{\pi}{3}$ from their
neighbors without defaults
(one site out of seven would not touch any dimer, see Fig.~\ref{klein})
and this {\em local} property cannot be propagated to the entire lattice.
We stress that this $\pm\frac{\pi}{3}$
dimer correlation cannot picture the entire wave function.
It would not account for the low first-neighbor ``dimer density'':
$<\SiSj>$ at distance $|i-j|=1$
is only very weakly antiferromagnetic ($<\SiSj>\simeq-0.07$)
and for $N=28$, $<d_{1,2}>$ is even lower than the
second-neighbor one $<d_{1,11}>$, see
Table \ref{correl}.). The information provided by the dimer structure
of Fig.~\ref{dimerb} is
that a low energy state cannot be achieved without a local
order with a complex geometry.

\begin{figure}
	\begin{center}
	\mbox{\psfig{figure=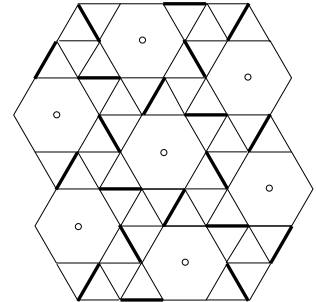,width=4cm}}
	\caption[99]{Dimer covering of a $6/7$-depleted triangular lattice.
	It is not possible to tile the triangular lattice with dimers
	forming angles $\pm \pi/3$ without defects. On site out of 7
	does not have any dimer (circles).
	It is worth noticing that a multiple-spin
	exchange Hamiltonian for which this state is an exact ground state can
	be constructed by the procedure introduced by Klein \cite{k82}.
	The Klein Hamiltonian is a sum of projectors: each projector acts on
	a rhombus and projects the four sites in their $S=2$ subspace.
	All the sites are equivalent on the depleted lattice and the
	Hamiltonian can
	be expressed as a Multiple-Spin Exchange
	Hamiltonian containing $P_{i,j}$
	at distance 1 ($J_2=5$) and $\sqrt{3}$ ($J_{2}^{\rm nn}=1$)
	and 4-body terms on rhombus ($J_4=1$ for $P_{1,2,3,4}+{\rm h.c.}$ and 
	$J_4'=2$ for a $P_{1,3}P_{2,4}$ term on the same rhombus).
	Actually, this state is also a ground state for the first-neighbor
	antiferromagnetic Heisenberg Hamiltonian on this lattice.}
	\label{klein}
	\end{center}
\end{figure}

\end{subsection}

\begin{subsection}{Chiral-chiral correlations:
	$<{\rm Im}(P_{i,j,k}) {\rm Im}(P_{i',j',k'}) >$}
	
	Kubo {\it et al.} have demonstrated that the classical ground state
	of the $J_4=1$ Hamiltonian was a 4-sublattice N{\'e}el state, where
	the magnetization vectors of the 4 sublattices form a regular
	tetrahedron \cite{km97,mkn97}. They showed with
	Monte-Carlo simulations that the {\em classical} system has
	a chiral phase transition at finite temperature\cite{mkn97}.
	This transition is associated with the ferromagnetic
	Ising-like order parameter defined on three sites by the operator:
	$\kappa=2(\vec{S}_1 \times \vec{S}_3)\cdot \vec{S}_2$.
	In the classical ground state all triangles
	have a chirality $\kappa=+\frac{\sqrt{3}}{9}$ or
	$\kappa=-\frac{\sqrt{3}}{9}$. Kubo's
	simulations indicate that the $<\kappa_{(1,2,3)} \kappa_{(1',2',3')}>$
	correlation function remains
	long-ranged up to a finite critical temperature.
	To check this hypothesis in the quantum case, we have computed this
	correlation in the ground state of the {\em quantum} system
	for the pure $J_4$ hamiltonian (like Kubo {\it et al.})
	and found a significant decay with distance
	(Fig.~\ref{chirality}). Moreover, this correlation shows
	negative values, which is unexpected for a ferromagnetic
	Ising model. The existence of such Ising LRO in the quantum
	system, even at T=0, seems unlikely.

\begin{figure}
	\begin{center}
	\mbox{\psfig{figure=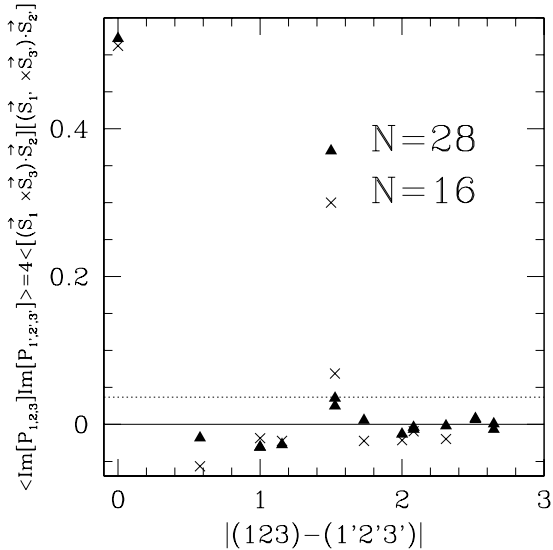,width=8cm}}
	\caption[99]{
	Chiral-chiral correlation function for the $J_4=1$ Hamiltonian.
	The distance is measured in lattice spacing units between the center
	of gravity of the two triangles.
	The chirality is simply expressed with spin exchange operator:
	$
		{\rm Im}(P_{i,j,k})=\frac{1}{2i}(P_{i,j,k}-P_{k,j,i})=
		2(\vec{S}_i \times \vec{S}_k)\cdot \vec{S}_j=\kappa
	$. 
	The value in the classical tetrahedral state is
	$<{\rm Im}[P]{\rm Im}[P']>=\frac{1}{27}\simeq 0.037$
	for any couple of clockwise triangles $P$ and $P'$
	(dotted line).
	If the two triangles have some sites in common,
	$<{\rm Im}[P]{\rm Im}[P']>$ may have a (small) imaginary part. In
	this case, the real part is plotted.}
	\end{center}
	\label{chirality}
\end{figure}
	
	It is true that the local fluctuations of $\kappa$ are large: for $N=28$
	($T=0$)
	$<\kappa^2>=0.5221$ which is larger that the expectation
	value on 3 {\em free} spins: ${\rm Tr}[\kappa^2]/{\rm Tr}[1]=0.375$.
	On the other
	hand, it is both much smaller than the maximum quantum value of $\kappa^2$
	($=\frac{3}{4}$) and much smaller than the value obtained in the low
	lying singlets on the Kagome lattice
	($<\kappa^2>\simeq 0.7$)~\cite{web98}.
	The last reason why the value of $<\kappa^2>$ should not be interpreted
	as sufficient evidence of $T$-symmetry breaking is that it contains
	no information but the first-neighbor correlation:
\begin{equation}
	<\kappa^2>=\frac{3}{2}\left(\frac{1}{4}-<\SiSj>\right)
	\label{kappa}
\end{equation}
	
	We also
	performed computations at the $J_2=-2$ and $J_4=1$ point.
	Unsurprisingly, the ferromagnetic coupling reduces the chirality.
	In the ground state of $N=28$ at $J_4=1$ and $J_2=-2$ we found
	$<\kappa^2>=0.4791$. This is readily understood from Eq.~\ref{kappa}.
	Since $<\SiSj>$ is enhanced by a negative $J_2$, local fluctuations
	of chirality are reduced. It is more interesting
	to look at $<{\rm Im}[P]{\rm Im}[P']>$ as a function of distance.
	The data are shown in Tab.~\ref{chirality2} and support an absence
	of long-ranged chiral order at the $J_2=-2$ and $J_4=1$ point
	as at the pure $J_4$ point.

\begin{table}
	\begin{tabular}{|c|c|c|}
	$|(123)-(1'2'3')|$ & $J_2=-2$ $J_4=1$ & $J_2=0$ $J_4=1$	\\ \hline
	1				& -0.0285	& $-0.0306$	\\
	$\frac{\sqrt{21}}{3}=0.527$	& +0.0280	& +0.0356	\\
	2				& -0.0078	& -0.0132	\\
	$\sqrt{7}=2.646$		& +0.0018	& +0.0009
	\end{tabular}
	\caption[99]{
	Chiral-chiral correlation in the ground state of an $N=28$ system.
	$<{\rm Im}(P_{1,2,3}) {\rm Im}(P_{1',2',3'})>$ at $J_4=1$ and for two
	values of $J_2$. $(1,2,3)$ and $(1'2'3')$ are clockwise triangles
	and $|(123)-(1'2'3')|$ is the distance between the
	centers of the two triangles. The increase of the chirality
	at distance $\sqrt{7}$ is very likely to be a finite size effect.
	For the three other distances we observe a significant decrease of
	chirality
	due to the first-neighbor ferromagnetic coupling ($J_2=-2$).
	}
	\label{chirality2} 
\end{table}

\end{subsection}

\end{section}

\begin{section}{Low energy spectral properties}

In the previous section we have reported some correlation
data showing that the ground state wave function of the MSE
Hamiltonian in the spin-liquid phase has short-ranged correlation
functions. We now describe the low energy spectrum of the system.
We emphasize on the spin gap and the broken symmetries.
Again, we concentrate on the $J_2=-2$, $J_4=1$ Hamiltonian.

\begin{subsection}{Spin gap}

\begin{subsubsection}{Finite size effects on the ground state energy}
For small sizes the spin gap and ground state energy are sensitive
to the sample shape
and size. For this reason, an $N=\infty$
extrapolation of the spin gap value is not straightforward
and requires a precise analysis.
The apparently irregular behavior
of the spin gap and ground state energy for systems with less than
$N\simeq24$ sites (Fig.~\ref{ega})
indicates the short-ranged rigidity of the ground state wave function.
This agrees with the previous section:
the characteristic length $\xi$ of the ground state correlations is a
few lattice sites and for systems of this size, the spectrum is
sensitive to boundary conditions. This effect
can be used to probe the local structure of the
ground state wave function and to identify some short wave
length resonances which lower its energy.
Two families
of samples should be distinguished: systems with even
$N$ have a low ground state energy and a higher spin gap
that the systems with $N$ odd. Within each family,
the behavior with the size is more or less monotonous:
for even-N systems (resp. odd-N) the energy per
site increases (resp. decreases) with $N$ and the spin gap decreases
(resp. increases) with $N$.
The importance of a
6-site periodicity for achieving a low energy\footnote{
	Systems with the lowest ground state energy per site
	turn out to be those with
	wave vectors
	${\bf k}={\bf A},{\bf B}$ and ${\bf B/2}$ in their
	first Brillouin zone.
	An illustration of this sensivity to the shape is
	provided by the two $N=24$ systems of Fig.~\ref{egb}.
	The more stable $N=24$
	system is the $6\times4$ torus (which possesses the
	four wave vectors mentioned and $E/N=-4.03$). It can be compared
	with the most compact 24-site torus
	(label 24' in Fig.~\ref{egb} and empty square
	at $N=24$ in Fig.~\ref{ega}).
	The later has only one ${\bf k}={\bf A}$ vector and is
	more frustrated: $E/N=-3.88$.
	The same behavior was observed for 12 sites.
	}
agrees with the $\pm \pi/3$ orientation
found by looking at the short-ranged dimer-dimer
correlations (Fig. \ref{dimerb}). The enhancement of this
$\pm \frac{\pi}{3}$ structure in the lowest energy systems
shows that it captures some important short distance
correlations.

An important point is that
both families, odd and even, merge for sizes $N\geq N_0\simeq 30$,
(either if we look at $E/N$ or
at the spin gap). We have interpreted this feature as a
crossover behavior: above the characteristic size $N_0\simeq \xi^2$
the finite size correction decay exponentially
fast\footnote{
	Notice that the situation would be different in the
	case of NLRO.
	The finite size corrections
	of the ground state energy per site are proportional to
	$N^{-3/2}$
	(linear spin-wave result).
	The energies of the different families of samples
	(those which frustrate the LRO and those which do not) would
	intercept linearly at $N=\infty$ in a plot of $E/N$ versus
	$N^{-3/2}$. As for the spin gap, it would vanish as $N^{-1}$:
	this is definitely inconsistent with present results
	(Fig.~\ref{ega})
}. Fig.~\ref{gapkA} also shows a rapid decay of finite size
effects when approaching $N=36$.
From this argument, the $N=36$ system can almost be considered
as having reached the thermodynamic limit. This is all the more
probable as this sample ($6\times6$) has the symmetries of the
infinite lattice and does not frustrate any short-ranged order
that we observed to be favorable.
\end{subsubsection}

\begin{subsubsection}{Spin gap}
The correlation
between the ground state energy and the energy gap to the first
$S=1$ (and $S=\frac{3}{2}$, if $N$ is odd) excited state is indeed precise
(see Fig.~\ref{egb}):
the most stable systems (i.e. with $E/N$ minimum) are those
with the largest spin gaps. The different spectra
roughly fall onto one line when the spin gap is plotted versus $E/N$.
This suggests that the position on this line is a single parameter
which measures the frustration of a system.
On the right part of Fig.~\ref{egb}, systems have important
frustration, because of geometric constraints the local structure
of the wave function which would minimize the energy in the infinite
systems cannot develop. Systems on the left part benefit of particular
boundary conditions which allow some stabilizing resonances
and therefore an $E/N$ lower than
in the thermodynamic limit.
From the data of $E/N$ versus $\Delta$
we can estimate the position of the
infinite system. $E/N$ converges to $-3.95 \pm 0.05$ (safe estimate
from Fig.~\ref{ega}). With this energy interval, 
the value of the spin gap can then be extracted from Fig.\ref{egb}.
We obtain $\Delta = 1.3 \pm 0.5$.
\end{subsubsection}
\begin{figure}
	\begin{center}
	\mbox{\psfig{figure=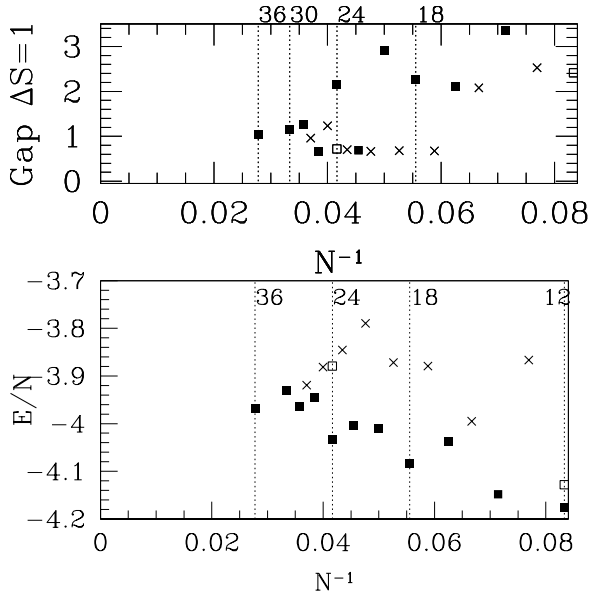,width=8cm}}
	\caption[99] {
	Energy and spin gap for $J_2=-2$ and $J_4=1$ as
	a function of $1/N$.
	{\bf Squares}: even $N$ systems. {\bf Crosses}: odd $N$.
	When different shapes have been investigated
	(which is the case for most
	sizes), only the shape with the lowest ground state energy
	was plotted.
	The exception for $N=24$ and $12$ is meant to illustrate
	the simultaneous
	decrease of the spin gap and increase of the ground state
	energy when the spins are arranged in a frustrating shape.}
	\label{ega}
	\end{center}
\end{figure}

\begin{figure}
	\begin{center}
	\mbox{\psfig{figure=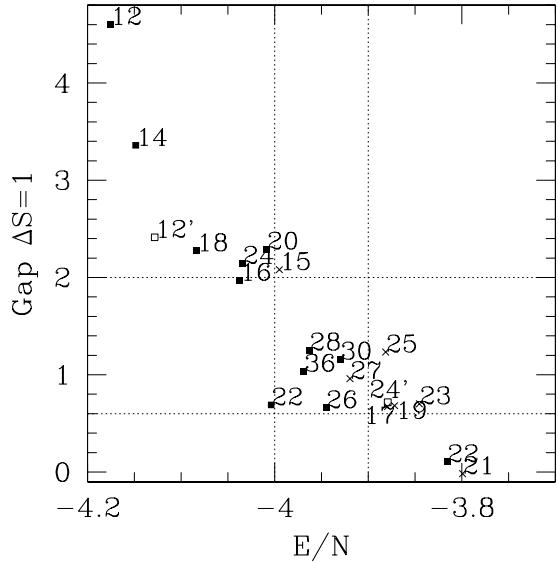,width=8cm}}
	\caption[99] {Same data as in Fig.~\ref{ega}
	with the spin gap plotted as a function of the ground state energy.}
	\label{egb}
	\end{center}
\end{figure}

\end{subsection}
\begin{subsection}{Ground state symmetries in the thermodynamic limit}
The spin gap described in the previous paragraph insures
that in this region of the phase diagram
the system does not spontaneously break 
the $SU(2)$ symmetry at low temperatures. We now turn to the
$S=0$ sector and address the question of the ground state
degeneracy
and the possibility for a spatial group symmetry breaking.
This is important to characterize the class of spin liquid
this model belongs to.

\begin{subsubsection}{Singlet states in the spin gap}
In nearly all even samples ($N=$36, 30, 28, 24, 18, 16, 12)
the ground state belongs to the
trivial representation of the point group and translation group.
This representation being one-dimensional,
the ground state is non-degenerate for finite $N$.
Above the ground state we find always less than
10 eigenstates in the spin gap ({\it i.e.} below the
first triplet energy). There is no extensive
entropy at $T=0$ nor any singlet soft mode (notice the difference
with the Kagome lattice Heisenberg antiferromagnet \cite{web98}).
This implies that the system at $J_2=-2$ and $J_4=1$ is
{\em not} a quantum critical system.

We have analyzed the (spatial) quantum numbers of the
low energy singlet states to detect if some
symmetry sectors were collapsing to the absolute ground state.
So far, because of the poor regularity in the symmetries,
this task has only brought partial results. Here are
the most important data to be understood:
\begin{itemize}
	\item There is no low-energy singlet in the
	${\bf k}={\bf B}$ sector
	($\pm{\bf B}$ are the two corners of the Brillouin zone). This excludes
	any 3-step translation symmetry breaking.
	It also confirms that the $\pm\pi/3$ dimer order
	(Fig. \ref{dimerb}), which looks 3-step periodic,
	is only {\em local}.
	\item The frequent occurrence
	of ${\bf k}={\bf A}$
	states in the spin gap suggests
	that such states collapse to the ground state and give
	a 2-step translation symmetry breaking
	(${\bf A}$ is one
	middle of the boundary of the Brillouin zone). In particular
	the energy gap between the absolute ground state and the
	first ${\bf k}={\bf A}$ state drops 
	by a factor of ten between $N=24$ and $N=36$,
	(2.5 at $N=24$ and 0.243 for $N=36$ see Fig~\ref{gapkA}).
	\item The $N=36$ singlet sector
	of the $J_2=-2$ and $J_4=1$ Hamiltonian
	has a quasi four-fold degeneracy of the ground state
	(Fig.~\ref{sp36}).
	The lowest energy is a single state and the next energy
	level, which is
	immediately above, is 3-fold degenerate.
	Because of the large size of the $N=36$ system (compared
	to the supposed correlation length), we believe that
	this feature is a particularly valuable information on the spectrum
	of the infinite system.
\end{itemize}	
In the next paragraphs we analyze these results.
First, we argue that these data are not in favor
of  ``spin-Peierls'' like symmetry breaking. Then we show
that a four-fold degeneracy could be of topological origin. Thirdly,
we explain why the system is probably not a Chiral Spin Liquid.
Eventually, we discuss the link between the MSE spin liquid and some
kind of AKLT (or Haldane) phase on a square superlattice with four spins
in the unit cell.
\begin{figure}
	\begin{center}
	\mbox{\psfig{figure=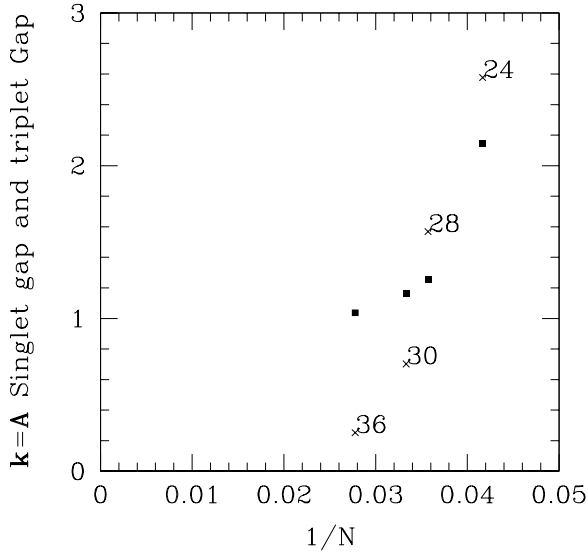,width=8cm}}
	\caption[99]{Spin gap (squares) and gap to the first
	{\bf k}={\bf A} singlet excited state (crosses).}
	\label{gapkA}
	\end{center}
\end{figure}
\end{subsubsection}
\begin{subsubsection}{No spin-Peierls symmetry breaking}
The $N=36$ singlet sector
of the $J_2=-2$ and $J_4=1$ Hamiltonian
has a quasi degeneracy of the ground state (Fig.~\ref{sp36}).
The lowest energy is a single state and the next one
is 3-fold degenerate
(impulsions ${\bf k}={\bf A}_{1,2,3}$).
As mentioned before,
the behavior of the singlet gap to the first ${\bf A}_{1,2,3}$
state (Fig. \ref{gapkA}) suggests that these two levels are degenerate
in the thermodynamic limit. The next pair of energies in $S=0$ may also
be seen as asymptotically degenerate
${\bf k}={\bf 0},{\bf k}={\bf A}_i$ states (Fig.~\ref{sp36}).

The point is that no pattern gives a 4 or 8-fold
degeneracy compatible with the quantum numbers
of these 4 or 8 lowest eigenstates. In particular, any
long-ranged ordered dimer
covering leads at least to 12 degenerate states. For
bigger patterns (4-site rhombus for instance) the
number of broken symmetries is even bigger and so
would be the degeneracy. Therefore, the 1+3 degeneracy
of the two first eigenvalues
(${\bf k}=0$ and ${\bf k}=A_{1,2,3}$) cannot be explained
in such a conventional spin-Peierls like picture.
\end{subsubsection}
\begin{subsubsection}{Topological degeneracy ?}
But this degeneracy
matches the prediction of a topological degeneracy
due to the periodic boundary conditions and to the
non-trivial topology of the torus.

Most of the important arguments for this topological degeneracy
were proposed by Rokhsar and Kivelson\cite{rk88},
Read and Chakraborty\cite{rc89}, 
and Sutherland\cite{s88} but here we sketch a more
formal demonstration.

{\it Dimer and loops}. ---
The starting point is a dimer representation
of the wave function. As remarked
by Sutherland\cite{s88}, the overlap
$<C|C'>$ between
two normalized valence-bond states $C$ and $C'$ can be diagrammatically
computed by constructing their {\em transition graph}.
This graph is obtained by drawing on the same lattice
the oriented bonds of $C$ and $C'$.
The result is a set of locally-oriented and non-intersecting loops
which cover the lattice.
Let $n(C\cup C')$ be the number of loops and
$x(C\cup C')$ the number of bonds
to be reversed for each loop to have alternating bond orientations.
The result is
\begin{equation}
	<C|C'>=2^{n(C\cup C')-N/2}(-1)^{x(C\cup C')}
\end{equation}
As an illustration, consider the trivial case $C=C'$.
All loops are of length 2 and
their number is $n(C\cup C)=N/2$. Each 2-site loop is already alternated
and $x(C\cup C)=0$. So we get $<C|C>=1$.
From the overcomplete family
of valence-bond states
one can formally extract a (non-orthogonal) basis.
In this basis, the ground sate reads:
\begin{equation}
	|\psi>=\sum_C \psi(C) |C>
\end{equation}
The normalization of $|\psi>$ gives
\begin{equation}
	<\psi|\psi>=\sum_{C,C'} \left[
		\psi(C)^\dagger\psi(C')2^{n(C\cup C')}(-1)^{x(C\cup C')}
		\right]
\end{equation}

{\it Small loops hypothesis}. ---
This equation can be re-written in term of a sum over
the transition graphs, or loop coverings $\mathcal G$ of the lattice:
\begin{equation}
	<\psi|\psi>=\sum_{\mathcal G}
		2^{n(\mathcal G )}(-1)^{x(\mathcal G)}
		\Gamma(\mathcal G)
	\label{graphs}
\end{equation}
where we defined:
\begin{equation}
	\Gamma(\mathcal G)=	\sum_{C,C'/C\cup C'=\mathcal G}\psi(C)^\dagger\psi(C')
	\label{gamma}
\end{equation}
This is possible because any decomposition $C\cup C'$
of a loop covering $\mathcal G$ has the same
$(-1)^{x(C\cup C')}=(-1)^{x(\mathcal G)}$ as
well as the same $n(C\cup C')=n(\mathcal G)$.
We now turn to
the important hypothesis:
we assume that the only graphs $\mathcal G$
which contribute to Eq.~(\ref{gamma}) are those which do not contain
any loop surrounding the torus.
This statement about the non-contribution of topologically non-trivial
loops implies that there should not be any long distance
coherence in the wave function. We stress that this is a much stronger
requirement that demanding short dimers (for instance, a
Valence Bond {\em Solid} or a
Valence Bond {\em Crystal} state does not fulfill this hypothesis).
In the MSE spin liquid all the correlations we measured
(spin-spin, chiral-chiral and dimer-dimer) seem to decay
exponentially over a distance $\xi$ and we do nothing but assume
that the loops contributing to Eq.~(\ref{graphs}) have
also a finite characteristic length-scale.
This strong condition implies the 4-fold degeneracy
of the ground state. We propose the arguments summarized below.

{\it Cuts along the torus and 4-fold degeneracy}. ---
The idea is now to chose a cut $\Delta$
surrounding the torus in one direction.
Each valence bond-state $C$ has a number
$\Delta(C)$ of bonds crossing $\Delta$ and it is
possible to change the sign of the configurations
$C$ for which $\Delta(C)$ is odd:
\begin{equation}
	|\psi'>=\sum_C (-1)^{\Delta(C)}\psi(C) |C>
\end{equation}
If the sample has an odd number of sites in the direction
perpendicular to $\Delta$ and if $|\psi>$ has an impulsion
${\bf k}$, it is easy to check that
the state $|\psi'>$ has an impulsion
${\bf k}'={\bf k}+{\bf A}$, where ${\bf A}$ is one
middle of the boundary of the Brillouin zone
(see inset of Fig.~\ref{sp36}).
But we have also shown\cite{m98} that $|\psi'>$ has the same energy
as $|\psi>$ provided  $|\psi>$
only ``contains'' small loops (hypothesis above).
This two-fold degeneracy must be extended to {\em four}
when the sample has a $2\pi/3$ rotation symmetry since
irreducible representations
for wave vector ${\bf k}={\bf A}$ are of dimension 3.
A similar construction was provided by Wen \cite{w91} in the
framework of a mean-field model for $P-$ and $T-$
symmetric spin liquid. In Wen's work, a state $|\psi'>$ is
built from the original ground state $|\psi>$ by a single gauge
transformation on the bonds degrees of freedom. This
transformation amounts to introduce a flux quantum through a non
contractable loop $\Delta$.

Now we can interpret the $N=36$ singlet sector
in the following way: as the system becomes larger
than $\xi$ the topologically non trivial loops
grow scarce exponentially with $\sqrt{N}/\xi$.
Consequently the lowest singlet eigenstate become
4-fold degenerate (${\bf k}=0$ and ${\bf k}=A_{1,2,3}$).
At $N=36$ this topological degeneracy is still not very
accurate
but we can clearly distinguish two quasi quadruplets. 
	
\begin{figure}
	\begin{center}
	\mbox{\psfig{figure=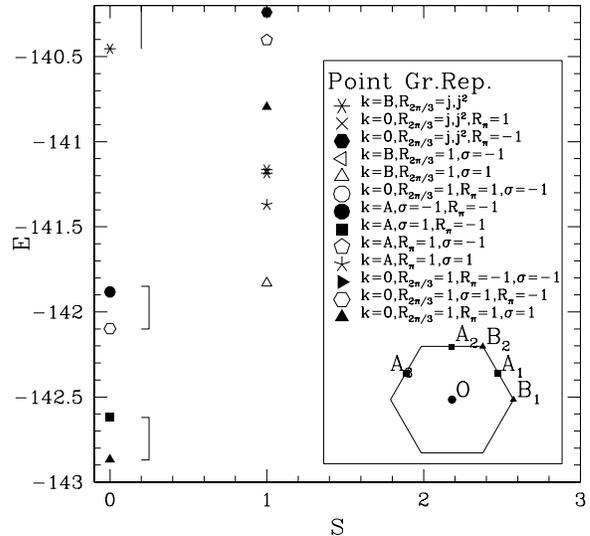,width=8cm}}
	\caption[99]{Lowest energy levels of the $N=36$ sample, ($J_2=-2$
	and $J_4=1$) in the sectors of ${\bf k}=0,{\bf A}$ and ${\bf B}$.
	The symbols represent the quantum number of the different
	eigenstates.
	For instance, the ground state ($S=0$, $E=-142.867$) belongs to
	the trivial representation
	of the spatial group (full upward triangle,
	labeled ${\bf k}=0,R_\frac{2\pi}{3}=1,R_\pi=1,\sigma=1$).}
	\label{sp36}
	\end{center}
\end{figure}
\end{subsubsection}

\begin{subsubsection}{A Chiral spin liquid ?}
Some time ago it has been suggested that an antiferromagnetic
quantum spin liquid could be {\em chiral} and could break the
time-reversal (T) and parity (P) symmetries\cite{kl87,kl89,wwz89}.
This scenario was definitely discarded
in the case of the simple Heisenberg Hamiltonian on the
triangular lattice\cite{blp92}.
The situation on the Kagome lattice is more ambiguous\cite{web98}.
In the MSE case, the presence of a gap in the magnetic excitations and
the degeneracy
of the ground state in the thermodynamic limit
would be consistent with a chiral theory.
Indeed, Wen \cite{w89} argue that an incompressible chiral fluid should
have a $2k^g$-fold vacuum degeneracy on a manifold of genius $g$
(but this is not specific to chiral models: one also finds
a $2^g$ topological degeneracy in $P$- and $T$-symmetric models
\cite{w91}).
On the other hand, as underlined before, the MSE does not support
LRO in the chiral parameter 
$\kappa=2(\vec{S}_1 \times \vec{S}_3)\cdot \vec{S}_2$. Therefore, the only possibility
would
be an hypothetical ``Chiral Spin Liquid'' ground state
\cite{kl89,wwz89} where the expectation
value of a cyclic permutation operator $P_{i_1,i_2,i_3,\ldots,i_n}$
acquires a non-zero imaginary part on large loops $i_1,i_2,i_3,\ldots,i_n$.
In such theories elementary excitations are unconfined spin-$\frac{1}{2}$
excitations. The spectra of $N=27$ and $N=28$,
are the largest pair of consecutive sizes we diagonalized and are
believed to be close to the asymptotic limit. Their comparison does
not plead in favor of this unconfined spinon hypothesis: the
ground state energy per site of $N=27$ ($E^{S=0}/N=-3.919$) is
slightly lower than the first
triplet energy per site of $N=28$ ($E^{S=1}/N=-3.918$ and
$E^{S=0}/N=-3.963$). This is
rather a signature of elementary {\em spin-1 excitations} and makes
the chiral scenario unlikely.

\end{subsubsection}
\begin{subsubsection}{An AKLT / Haldane system ?}

We have shown, so far, that the MSE spin liquid is characterized
by a low ground state degeneracy. For half-integer spins in one dimension,
the LSM\cite{lsm61} theorem forces the ground state to be degenerate
or gapless.
Some attempts to generalize it in two dimensions\cite{a88}
suggests that a ground state degeneracy could also be generic in
a (half-integer) gapped state in two dimensions. From this point
of view, our results in the MSE case are not unexpected.
But such a degeneracy is usually  associated to
a spontaneous breaking of a discrete symmetry, say translation, and
a dimer or plaquette local order parameter.
If our analysis of the $N=36$ spectrum is correct, such a spin-Peierls
phenomenon is absent.
Without any evidence of a simple local order parameter,
the topological point of view is natural and  
states that a ground state degeneracy can occur
without any local order parameter breaking a discrete symmetry
(a famous example is the Laughlin wave function, which discrete
degeneracy depends on the space topology \cite{wn90}).
However, the translation symmetry is almost
certainly broken in the MSE case (at least four 
states with different impulsions are asymptotically degenerate)
and one should be careful before excluding the possibility of {\em
any} kind of LRO.
In search for an alternative to the topological degeneracy,
we looked  for explicit
wave functions (for half-integer spin) with a low
degeneracy and the minimum number of broken spatial symmetries
to compare with numerical results.

For high spins matching
the coordination number ($2S=z$) the AKLT wave function
is non degenerate.
When this criterion is not satisfied, nevertheless, the AKLT procedure
can help
to write wave functions with few broken symmetries.
The idea is to chose a superlattice with $n$ spins in the unit cell and of
coordination number $z=n$. On the superlattice, a unique AKLT state is
written by
identifying the $n$ spins of a cell as a larger spin $S=\frac{n}{2}$.
The state we obtained has a degeneracy which is the number of possibilities
to define a superlattice $L$ over the real (triangular) lattice.
Consider the $S=2$ AKLT wave function for the square lattice. It is mapped
onto the spin$-\frac{1}{2}$ triangular lattice
by identifying the $S=2$ spins of the
square lattice to 4 spin$-\frac{1}{2}$ (on a rhombus) of the triangular one
(Fig.~\ref{AKLT}.a).
The choice of the position of the square superlattice
is done among twelve different possibilities: a configuration can be
translated at four different places and rotated in three directions.
One should remark that these states do have some long-ranged order.
Let $\Pi_a$ be the projection operator of the rhombus $a$ in its
$S=2$ subspace. By construction, 
the 8-point correlation function $<\Pi_a\Pi_b>$ is long-ranged.

We computed the ground state energy inside this 12-dimensional AKLT subspace
on a small system ($N=12$ at $J_4=1$ and $J_2=-2$)
and found a good variational energy.
The exact ground state energy
is $E/N\simeq-4.128$ and the AKLT energy is $E/N=-3.75$.
The energy scale is given by the energy of the ferromagnetic state ($E/N=0$),
and the energy of the ground state $E/N=-3$ for classical spins.
This variational result is {\em the best achieved so far} and is
even much better than the Schwinger-boson
energy for this Hamiltonian (the Schwinger-boson solution
is a two-sublattice (NLRO) collinear state with $E/N=-2.96$).
The reason for this competitive energy is
that the AKLT construction naturally provides
some local ferromagnetism inside a plaquette which lowers $J_2 P_{1,2}$
and some antiferromagnetism at larger distance (between neighboring rhombus)
which insures a low $J_4\left(P_{1,2,3,4}+{\rm h.c}\right)$.
When the size of
the unit cell is varied, one tunes the strength of the antiferromagnetic
correlations at short distance. For instance, it is
possible to enhance it if we start from the wave function
of the $S=3/2$ honeycomb lattice (Fig.~\ref{AKLT}.b), where the unit cell
has only 3 sites. In this case we find an energy slightly higher
($E/N=-3.42$ on 12 sites). On the other hand, varying the ferromagnetic
coupling $J_2$ in the Hamiltonian could bring the ground state closer
to an AKLT wave function.

The twelve AKLT states are not coupled by any local Hamiltonian in
the thermodynamic limit, and should therefore be degenerate in an
infinite system. 
In a theory where all excitations are gapped, the low energy
physics (except for possible edge states) is determined by the degenerate
ground state. Therefore, this twelve-fold degeneracy characterizes
the fixed point
and we expect the same degeneracy in the MSE problem if it
belongs to this universality class.
Unfortunately the quantum numbers only partially coincide with
the lowest eigenstates found in the $N=36$ spectrum. So, despite of
a good variational energy on $N=12$, these particular trial states do
not provide a quantitative explanation of the 4 or 8-fold degeneracy
observed on 36 sites. However, the family of these extended AKLT states
certainly captures a part of the local constraints imposed
by frustrating MSE couplings. Furthermore we cannot {\em completely} exclude
that the $N=36$ might not be a faithful picture of the infinite system
spectrum and that the ground state could have a complex
Valence Bond Solid order
of the type we have just described.

\begin{figure}
	\begin{center}
	\mbox{\psfig{figure=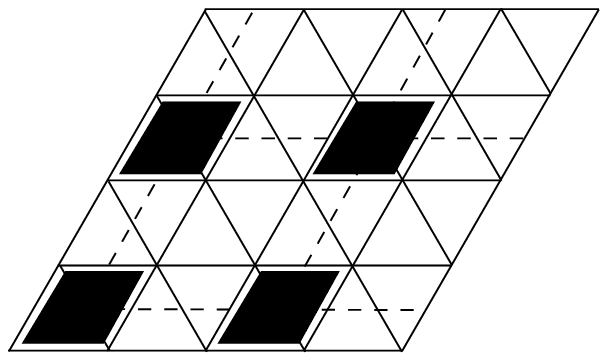 ,width=4.1cm}}
	\mbox{\psfig{figure=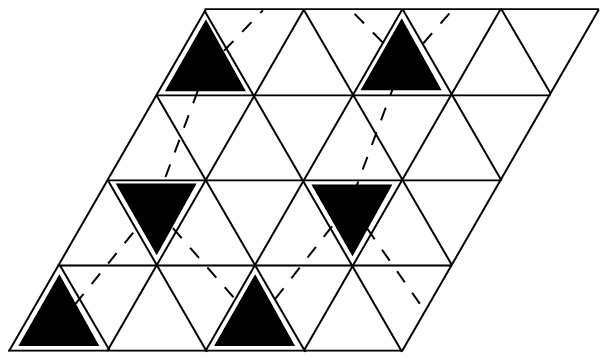,width=4.1cm}}
	\end{center}
	\caption[99]{{\bf Left}: black rhombus represent the $S=2$ spins of
	the AKLT wave function. They form a square lattice (dashed lines).
	{\bf Right}: honeycomb superlattice of 3-site plaquettes
	on the triangular lattice. There are 18 different ways to place
	this honeycomb superlattice on the triangular one ($3\times2$ translations
	and 3 rotations).}
	\label{AKLT}
\end{figure}

\end{subsubsection}
\end{subsection}

\end{section}

\begin{section}{Summary and Conclusions}

We analyzed the spin liquid phase of MSE model on the triangular lattice.
Due to important frustration, LRO is destroyed by quantum fluctuations
at zero temperature. The system has short-ranged
correlations: $<\SiSj>$, dimer-dimer and
chiral-chiral correlations probably decay exponentially with distance.
The spectrum has a spin gap and the comparison between
odd and even sample pleads in favor of spin-1 excitations.
As for the spatial symmetries and degeneracy
of the ground state (presumably 4 or 8), they provide evidence of
no spin-Peierls nor simple
plaquette order. This is a generic feature in systems
where the spin matches the coordination number
($2S=z$) and where an AKLT wave function can be constructed,
but it is unconventional for spin-$\frac{1}{2}$. 
We proposed two interpretations.
The first possibility is an AKLT or Haldane -like phase.
We constructed such a wave function,
which is not a tensor product of plaquette states. Present numerical
data on the ground state symmetries do not completely agree with
this particular trial state
but, because of its very low variational energy,
such a scenario cannot be excluded.
Secondly, this degeneracy might be a consequence of the
non trivial topology implied by periodic boundary conditions.
We emphasized that it must be the case if the system has absolutely
no long-ranged coherence.

This spin liquid phase might be the very first magnet with a
``disordered'' and gapped quantum ground state
with no simple local order parameter
in two dimensions. From this point of view
it can be seen as a prototype of the RVB state as was proposed
by Anderson\cite{a73}.

\end{section}

\end{document}